**REVIEW ARTICLE**

**Open Access**

# Competing few-body correlations in ultracold Fermi polarons

Ruijin Liu[1] and Xiaoling Cui[2*]

**Abstract**

Polaron, a typical quasi-particle that describes a single impurity dressed with surrounding environment, serves as an ideal platform for bridging few- and many-body physics. In particular, different few-body correlations can compete with each other and lead to many intriguing phenomena. In this work, we review the recent progresses made in understanding few-body correlation effects in attractive Fermi polarons of ultracold gases. By adopting a unified variational ansatz that incorporates different few-body correlations in a single framework, we will discuss their competing effects in Fermi polarons when the impurity and majority fermions have the same or different masses. For the equal-mass case, we review the nature of polaron-molecule transition that is driven by two-body correlations, and especially highlight the finite momentum character and huge degeneracy of molecule states. For the mass-imbalanced case, we focus on the smooth crossover between polaron and various dressed clusters that originate from high-order correlations. These competing few-body correlations reviewed in Fermi polarons suggest a variety of exotic new phases in the corresponding many-body system of Fermi-Fermi mixtures.

## 1 Introduction

In 1933, Landau raised the concept of polaron to describe electrons moving in solid materials while dressing with lattice distortions [1]. This concept depicts a fundamental quasi-particle picture and, nowadays, has been successfully applied to various physical systems such as high-Tc superconductors [2], semiconductors [3], and neutral stars [4]. In recent years, the study of polaron physics has gained significant developments in ultracold atoms [5, 6], thanks to the high controllability there over atomic species, number, and interaction strength. Up to date, both Fermi polaron [7–17] and Bose polaron [16, 18–23] have been successfully realized in ultracold mixtures, where the majority atoms are respectively with fermionic and bosonic statistics. Generally, these polaron systems exhibit two spectral branches in the radio-frequency spectroscopy, namely, the attractive lower and repulsive higher branches. These branches directly manifest the two-body correlation effect between impurity and majorities due to attractive or repulsive interactions. Beyond two-body correlations, theoretical studies have also revealed intriguing three-body correlations in both Fermi polarons [24–36] and Bose polarons [37–43]. Nevertheless, the experimental exploration of high-order correlations is still on the way.

Among various polaron systems, the attractive Fermi polaron emerges as a convenient testbed for emergent few-body correlations in a many-body environment. In this system, by tuning the mass ratio between the majorities and the impurity, different few-body correlations can become dominant and lead to distinct phenomena of Fermi polarons, see illustration in Fig. 1. For instance, for the equal-mass Fermi polaron, the enhanced two-body correlation as increasing the impurity-fermion attraction will lead to a first-order polaron-to-molecule transition [44–52]. Such a transition has been extracted from various methods, including variational and quantum Monte-Carlo methods, based on a separate treatment of polaron

*Correspondence:
Xiaoling Cui
xlcui@iphy.ac.cn
[1] Institute of Theoretical Physics, University of Science and Technology Beijing, Beijing 100083, China
[2] Beijing National Laboratory for Condensed Matter Physics, Institute of Physics, Chinese Academy of Sciences, Beijing 100190, China

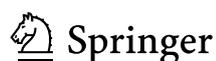

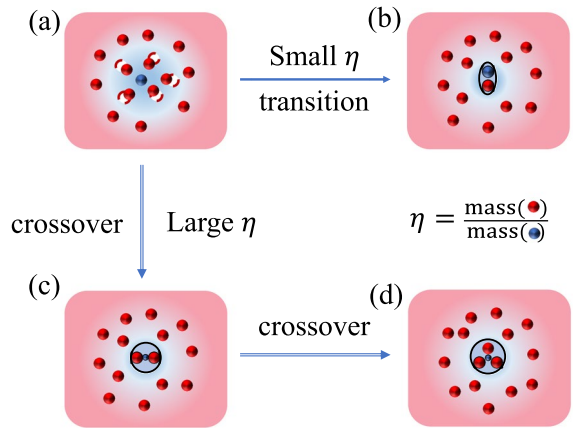

**Fig. 1** Different phases of attractive Fermi polarons. **a** Polaronic state, where the impurity attracts many fermions around and forms a quasi-particle. **b**, **c**, and **d** are respectively the dressed molecule, trimer, and tetramer states, where the impurity essentially binds with one, two, and three fermions to form few-body clusters on top of the majority Fermi sea. For small mass ratios between fermions and the impurity, the system undergoes a first-order transition from polaron (**a**) to molecule (**b**) states as the fermion-impurity attraction increases. In contrast, for sufficiently large fermion-impurity mass ratios, increasing fermion-impurity attraction will drive a smooth crossover of the system from polaron (**a**) to dressed trimer (**c**) and further to tetramer (**d**) states

and molecule states. In particular, the variational approach can be traced back to the famous ansatz proposed by Chevy [53], which method turns out to be very successful in describing the normal state of highly-polarized spin-1/2 Fermi gas. Later, by generalizing the Chevy's ansatz to finite momenta, we have shown that the polaron and molecule states can be unified into a single framework, and the nature of their first-order transition lies in the energy competition between different total momentum sectors [54, 55]. Under such a unified treatment, one can extract the coexistence regime of polaron and molecule directly from the energy dispersion of impurity system, and moreover, this approach reveals the hidden degeneracy of molecule states, which is important for estimating the actual polaron/molecule fraction near their coexistence regime.

Recently, the mass-imbalanced Fermi-Fermi mixture has become available in cold atoms experiments, such as $^{40}$K-$^{6}$Li [56–58], $^{161}$Dy-$^{40}$K [59, 60], $^{53}$Cr-$^{6}$Li [61–64], and $^{53}$Yb-$^{6}$Li [65, 66]. Compared to the equal mass case, the mass-imbalanced Fermi polaron turns out to be even intriguing due to the existence of few-body clusters therein. In the few-body sector, it has been shown that a light atom can bind with $N(\geq 2)$ heavy fermions as long as the heavy-light mass ratio is beyond certain critical value [67–78]. Physically, these bound states emerge because of the light-atom-induced effective long-range attraction between heavy fermions [68, 79]. Depending on whether the binding energies rely on short-range details or not, these cluster bound states can be classified as the Efimov [67, 69, 71] or universal [68, 70–78] types. In particular, the universal clusters are believed to be stable against inelastic collisions since the short-range details are irrelevant there, and their emergences have been shown to be very robust against the effect of finite effective range in quasi-two-dimension [80]. Given these mass-imbalance-facilitated few-body clusters, an important question is how their associated few-body correlations affect the Fermi polaron system. To this question, earlier studies have pointed out the dominant three-body correlation can drive the polaron-trimer transition [24, 25, 31, 35], again based on a separate treatment of polaron and trimer states. In contrast, by utilizing a unified variational ansatz that includes high-order particle-hole excitations (incorporating all essential few-body correlations on an equal footing), we have shown recently that instead of sharp transitions, the mass-imbalanced Fermi polaron undergoes a sequence of smooth crossover from polaronic state to various dressed cluster states as the impurity-fermion attraction increases [36]. This is distinct from the polaron-molecule transition as in mass-imbalanced case, and thus directly manifests the unique effect of high-order correlations in Fermi polaron systems. Recently, the variational approach has also been extended to finite temperatures and to cases with arbitrarily high-order particle-hole excitations [81, 82].

The organization of the rest of this paper is as follows. In Section 2, we introduce the unified variational approach for Fermi polaron problem, and further discuss its difference and relation to other variational ansatzs in literature. In Section 3, we review the properties of polaron-molecule transition and coexistence in equal-mass Fermi polarons, where special emphasis is put on the finite-momentum character and hidden degeneracy of molecule state. In Section 4, we turn to the mass-imbalanced case and point out distinct physical consequences of Fermi polarons under high-order correlations. Finally, we summarize in Section 5 and meanwhile comment on fascinatingly new phases in the many-body system of Fermi-Fermi mixtures due to competing few-body correlations.

## 2 Unified variational approach for Fermi polaron problem

We write down the Hamiltonian of Fermi polaron system ($\hbar = 1$):

$$H = \sum_{\mathbf{k}} \left( \epsilon^a_{\mathbf{k}} a^\dagger_{\mathbf{k}} a_{\mathbf{k}} + \epsilon^f_{\mathbf{k}} f^\dagger_{\mathbf{k}} f_{\mathbf{k}} \right) + \frac{g}{L^D} \sum_{\mathbf{Q},\mathbf{k},\mathbf{k}'} a^\dagger_{\mathbf{Q}-\mathbf{k}} f^\dagger_{\mathbf{k}} f_{\mathbf{k}'} a_{\mathbf{Q}-\mathbf{k}'}. \quad (1)$$

Here, $a^\dagger_{\mathbf{k}}$ and $f^\dagger_{\mathbf{k}}$ respectively create an impurity atom and a majority fermion with momentum $\mathbf{k}$ and energy $\epsilon^{a,f}_{\mathbf{k}} = \mathbf{k}^2/(2m_{a,f})$; $g$ is the bare coupling strength in $D$-dimension, and $L$ is the system size along one direction; Further, we define $k_F$ as the Fermi momentum of majority Fermi sea and $E_F = k_F^2/(2m_f)$ as the Fermi energy. The mass ratio between majority fermions and the impurity is introduced as $\eta \equiv m_f/m_a$.

Early in 2006, Chevy proposed a variational ansatz for the attractive Fermi polaron [53]:

$$|P(0)\rangle = \left( \psi_0 a_\mathbf{0}^\dagger + \sum_{\mathbf{kq}} \psi_{\mathbf{kq}} a_{\mathbf{q-k}}^\dagger f_\mathbf{k}^\dagger f_\mathbf{q} \right) |\text{FS}\rangle_N, \quad (2)$$

where $|\text{FS}\rangle_N$ denotes Fermi sea consisting of $N$ heavy fermions and all $\mathbf{k}$ ($\mathbf{q}$) stay above (below) the Fermi surface. This state describes a polaronic state of a zero-momentum impurity dressed with the lowest-order particle-hole excitations of the majority Fermi sea, which has been widely explored in literature. Later, this state was generalized to include two particle-hole excitations [25, 55, 83] for a more accurate description of attractive Fermi polaron, and even to describe the repulsive branch of Fermi polarons [84]. To compete with this polaronic state, a different variational ansatz was proposed in 2009 as [46]:

$$|M(0)\rangle = \left( \sum_\mathbf{k} a_\mathbf{k}^\dagger f_{-\mathbf{k}}^\dagger + \sum_{\mathbf{kk'q}} \psi_{\mathbf{kk'q}} a_{\mathbf{q-k-k'}}^\dagger f_\mathbf{k}^\dagger f_{\mathbf{k'}}^\dagger f_\mathbf{q} \right) |\text{FS}\rangle_{N-1}. \quad (3)$$

This is a molecule on top of the Fermi sea, carrying zero center-of-mass momentum and meanwhile dressed with particle-hole excitations. Then, by comparing the energies of polaronic and molecule states, a first-order polaron-molecule transition was concluded in 3D [46–49] and 2D [50] as increasing the impurity-fermion attraction. The same conclusion was also drawn from the Monte-Carlo method [44, 51, 52], where extracts the energies of polaron and molecule from different physical quantities. Similarly, one can construct a dressed trimer state when the three-body correlation dominates, and a polaron to trimer transition can be concluded by comparing the energies of these different states [24, 25, 31, 35].

Although Eqs. (2) and (3) have successfully captured different characters of polaron and molecule, the separate treatment of these states makes the first-order transition between them too artificial. To resolve this problem, a unified framework for different states is required. In Ref. [36], a unified ansatz is proposed for all relevant impurity states (polaron, molecule, trimer, etc.):

$$|P(\mathbf{Q})\rangle = \left( \psi_0 a_\mathbf{Q}^\dagger + \sum_{\mathbf{kq}} \psi_{\mathbf{kq}} a_{\mathbf{Q+q-k}}^\dagger f_\mathbf{k}^\dagger f_\mathbf{q} + \sum_{\mathbf{kk'qq'}} \psi_{\mathbf{kk'qq'}} a_{\mathbf{Q+q+q'-k-k'}}^\dagger f_\mathbf{k}^\dagger f_{\mathbf{k'}}^\dagger f_\mathbf{q} f_{\mathbf{q'}} + \ldots \right) |\text{FS}\rangle_N. \quad (4)$$

This is the direct extension of Chevy's ansatz in (2) to arbitrary momentum $\mathbf{Q}$ and to include high-order particle-hole excitations. We emphasize here that the extension to finite $\mathbf{Q}$ allows the description of molecule state, while the inclusion of high-order p-h excitations allows the incorporation of high-order correlations beyond the two-body one. To be specific, when $\mathbf{Q} = \mathbf{k}_F$, $|P_\mathbf{Q}\rangle$ well covers the molecule state $M(0)$ in (3). This can be seen straightforwardly by setting the hole momentum $\mathbf{q} = -\mathbf{k}_F$ in the second and third terms of (4), which directly reproduce the two terms in (3). This means that $M(0)$ in (3) only expands a subspace of all particle-hole excitations included in $|P(\mathbf{k}_F)\rangle$. As a result, $M(0)$ always has a higher energy than $|P(\mathbf{k}_F)\rangle$, and the former can only become a good approximation of the latter in strong coupling regime [54, 55, 85]. This is why $|P(\mathbf{Q})\rangle$ can serve as a unified ansatz for both polaron and molecule, see more discussions in Section 3.

Moreover, $|P(\mathbf{Q})\rangle$ can also incorporate high-order correlations by controlling the truncated level of particle-hole excitations. Because of much larger p-h excitation space, it is more energetically favorable than the dressed cluster states as proposed previously. In fact, it is noted that the $n$-body correlation effect in the system has been fully incorporated in the $(n-1)$th order particle-hole excitations in (4). In this way, one can consider various essential few-body correlations on the equal footing by controlling the truncated level of p-h excitations in (4). We shall discuss this in more detail in Section 4.

## 3 Polaron-molecule transition and coexistence in equal-mass Fermi polarons

In this section, we revisit the polaron-molecule transition in equal-mass Fermi polarons based on the unified ansatz in (4). The relevant results were first presented in Ref. [54] for 3D system based on (4) with the lowest p-h excitations, and then in Ref. [55] for both 3D and 2D systems up to two p-h excitations, which gives more accurate results for the location of polaron-molecule transition and coexistence. In the following, we shall focus on the 3D case.

To quickly identify different variational ansatz with given truncated level of p-h excitations, we rewrite (4) as

$$P_{2n+1}(\mathbf{Q}) = \left( \psi^{(0)} a_\mathbf{Q}^\dagger + \sum_{l=1}^n \sum_{\mathbf{k}_i \mathbf{q}_j} \psi^{(l)}_{\mathbf{k}_i \mathbf{q}_j} a_\mathbf{P}^\dagger \prod_{i=1}^l f_{\mathbf{k}_i}^\dagger \prod_{j=1}^l f_{\mathbf{q}_j} \right) |\text{FS}\rangle_N. \quad (5)$$

with $n$ the truncation number of p-h excitations. In this way, $P_{2n+1}(\mathbf{Q})$ has included all the two-, three-, ..., and $(n+1)$-body correlations in a single framework.

Similarly, we can write down the molecule ansatz up to $n$ p-h excitations:

$$M_{2n+2}(\mathbf{Q}_M) = \left( \sum_{\mathbf{k}} \phi_{\mathbf{k}} a^\dagger_{\mathbf{Q}_M-\mathbf{k}} f^\dagger_{\mathbf{k}} + \sum_{l=1}^{n} \sum_{\mathbf{k}_i \mathbf{q}_j} \phi^{(l)}_{\mathbf{k}'\mathbf{k}_i \mathbf{q}_j} a^\dagger_{\mathbf{P}} f^\dagger_{\mathbf{k}'} \prod_{i=1}^{l} f^\dagger_{\mathbf{k}_i} \prod_{j=1}^{l} f_{\mathbf{q}_j} \right) |FS\rangle_{N-1}. \quad (6)$$

In the following, since $\mathbf{Q}$ and $\mathbf{Q}_M$ are all rotational invariant, we shall use their amplitudes, $Q = |\mathbf{Q}|$ and $Q_M = |\mathbf{Q}_M|$, in $P_{2n+1}(Q)$ and $M_{2n+2}(Q_M)$ for simplicity.

Figure 2 shows the energy comparison of various ansatz for 3D Fermi polarons as the coupling strength changes. One can see that the molecule state $M_4(0)$ (or $M_2(0)$) always has a higher energy than $P_5(k_F)$ (or $P_3(k_F)$), and the former only has a good approximation of the latter when the system enters strong coupling regime. At $1/k_F a_s = 0.91$, the energies of $P_5(0)$ and $P_5(k_F)$ (well approximated by $M_4(0)$ at this coupling strength) cross each other, signifying the polaron to molecule transition at this point. This critical point is very close to the values obtained from Monte-Carlo [44] and diagrammatic [45] methods.

The nature of polaron-molecule transition can be seen clearly from the energy dispersion of $P_5(Q)$, as shown in Fig. 3 for various coupling strengths from weak to strong. In weak coupling regime, the dispersion shows a single minimum at $Q = 0$, denoting a polaronic ground state. Continuously increasing attractions, another (metastable) minimum appears at $Q = k_F$, and the two minima reach the same energy at the critical coupling $[1/(k_F a_s)]_c = 0.91$. In this sense, the nature of polaron-molecule transition lies in the energy competition between different momentum states $Q = 0$ and $Q = k_F$.

Above interpretation of polaron-molecule transition tells us two important facts that have been overlooked in previous studies. One is that the polaron and molecule can only exist when there are two locally stable minima in the energy dispersion, i.e., for $1/(k_F a_s) \in (0.5, 1.2)$ as shown in Fig. 3, but not for all coupling strengths. Their coexistence in realistic systems also requires a finite impurity density or a finite temperature, as pointed out in [54, 55, 86]. Second, the molecule states have a huge hidden degeneracy due to its associated finite momentum $|\mathbf{Q}| = k_F$. Given the rotational symmetry of $\mathbf{Q}$, the degeneracy manifold would be $SO(3)$ in 3D and $SO(2)$ in 2D. Physically, such degeneracy comes from the fact that the impurity can pick up any fermion at the Fermi surface to form the molecule, as shown schematically in Fig. 4. In this sense, the molecule ansatz $M(0)$ represents a symmetry-breaking state within the degenerate manifold. The large degeneracy leads to the significant enhancement of the density of state(DoS) of molecules in low-energy limit. This is because near $|\mathbf{Q}| \sim k_F$, the molecule dispersion is

$$E(\mathbf{Q}) = \epsilon_M + \frac{(|\mathbf{Q}| - k_F)^2}{2m^\star_M}. \quad (7)$$

From this dispersion, one can see that the DoS scales as $(\Delta E)^{-1/2}$ near the energy minimum ($\Delta E \to 0$), which is identical to the 1D system and diverges as $\Delta E \to 0$. As pointed out in Ref. [55], such significantly enhanced DoS can greatly favor the molecule occupation in the regime of polaron-molecule coexistence. This is to be confirmed in future unbiased experiment on Fermi polarons.

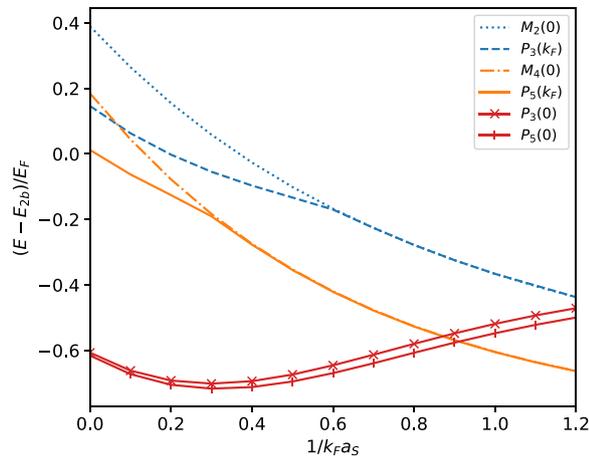

**Fig. 2** Energy comparison between various ansatz for 3D Fermi polaron system. All energies are shifted by $E_{2b} = -1/(ma_s^2)$ in $a_s > 0$ side in order to highlight the difference. Adapted from Ref. [55]

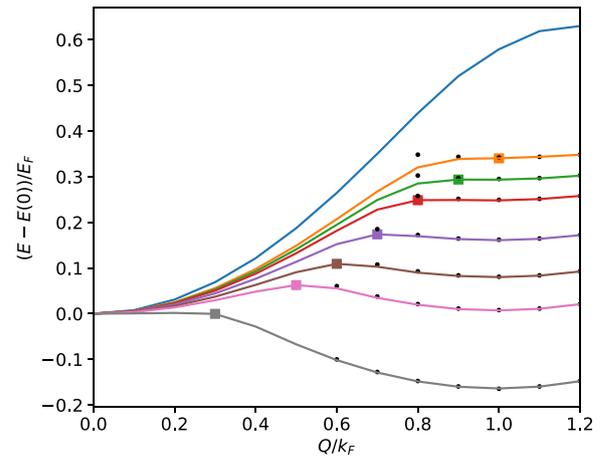

**Fig. 3** Energy dispersion of $P_5(\mathbf{Q})$ (solid lines) in 3D at various couplings (from top to bottom) $1/(k_F a_s) = 0.2, 0.5, 0.55, 0.6, 0.7, 0.8, 0.9, 1.2$, shifted by the value at $\mathbf{Q} = 0$. The rectangular point mark the position of maximum energy, and the small black dots show the energies of $M_4(\mathbf{Q}_M)$, with $|\mathbf{Q}_M|$ shifted by $k_F$ in order to compare with the energies of $P_5(\mathbf{Q})$. Here $Q = |\mathbf{Q}|$. Adapted from Ref. [55]

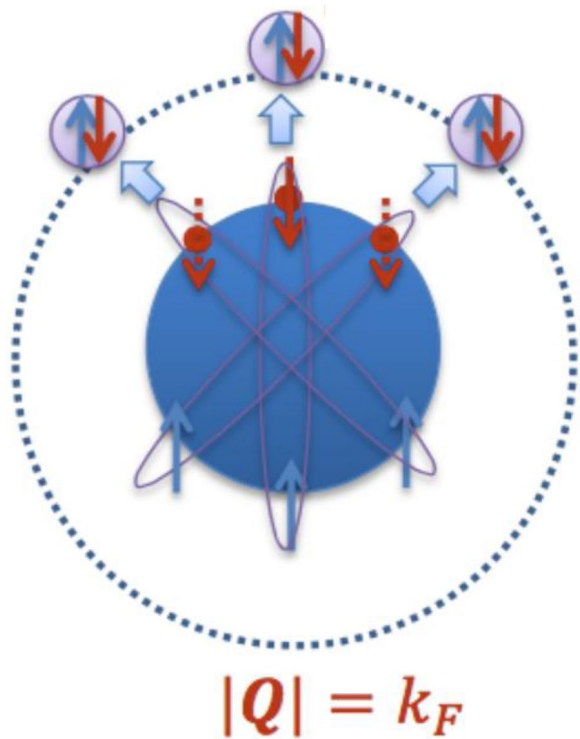

**Fig. 4** Illustration of the molecule degeneracy. In order to create a molecule outside the Fermi sea with zero center-of-mass momentum $\mathbf{Q}_M = 0$, the impurity has to be associated with momentum $|\mathbf{Q}| = k_F$, such that it can pair with a fermion at the Fermi surface to form the molecule with $\mathbf{Q}_M = 0$. The large degeneracy of molecule state comes from the fact that the impurity can pick up *any* fermion at the Fermi surface to form the molecule, and different configurations decouple with each other due to the conservation of total momentum $\mathbf{Q}$

Finally, we remark that the degeneracy of molecule state, not showing up in vacuum, is supported by the presence of majority Fermi sea ($k_F \neq 0$). It follows that the enhanced molecule DoS, as compared to vacuum case, also closely relies on the finite Fermi surface. This is very similar to the Cooper instability for two atoms on top of a spin-1/2 Fermi sea. In this latter case, the presence of Fermi sea also enhances the DoS of two scattering atoms (from 3D to 2D-like), such that an infinitesimal attraction among them can afford a molecule (pair) outside the Fermi sea. Here for the impurity case, the enhancement of DoS is even more dramatic (from 3D to 1D-like) given the dispersion in Eq. 7.

## 4 High-order correlations and smooth crossover in mass-imbalanced Fermi polarons

The mass-imbalanced Fermi polaron holds distinct properties from the equal mass case due to the presence of high-order correlations therein. This can be seen easily from the exact few-body solutions, which tell that a light atom can bind with a number of heavy fermions to form cluster bound state as long as the fermion-impurity mass ratio exceeds certain value [67–77]. Their binding mechanism can be understood under the Born-Oppenheimer approximation [68, 79]. Namely, the movement of the light atom induces an effective long-range attraction between heavy fermions, and when such attraction overcomes the p-wave centrifugal barrier between the fermions, they will bind together to former a stable bound state. Depending on the actual mass ratio, these bound states can be classified into the Efimov [67, 69, 71] and the universal [68, 70–77] types. The universal clusters do not depend on any short-range details of interaction potential, and thus they are believed to be stable against elastic atom loss. Importantly, their formation indicates high-order correlations in mass-imbalanced system that are well beyond the two-body ones. Given the growing number of mass-imbalanced Fermi-Fermi mixtures now available in cold atoms, such as $^{40}$K-$^{6}$Li [56–58], $^{161}$Dy-$^{40}$K [59, 60], $^{53}$Cr-$^{6}$Li [61–64], it is demanding to explore the effects of these high-order correlations in corresponding many-body systems.

In Ref. [36], we have studied the consequence of high-order few-body correlations in 2D Fermi polarons using the unified ansatz (Eq. 5) up to three particle-hole excitations, which incorporates the two-, three-, and four-body correlations in a single framework. The resulted phase diagram in the parameter plane of interaction strength ($= \ln(k_F a_{2D})$) and mass ratio ($\eta \equiv m_f/m_a$) is shown in Fig. 5. One can see that for small mass imbalance $\eta < \eta_{tr}$ (here $\eta_{tr}$ is the critical mass ratio to support a 2D trimer in vacuum), the system will undergo a first-order transition from polaron (with total momentum $\mathbf{Q} = 0$) to molecule (with $|\mathbf{Q}| = k_F$) as the impurity-fermion attraction increases. This is very similar to the equal mass case discussed in previous section. In contrast, for sufficient mass imbalance $\eta > \eta_{tr}$, the system shows no sharp transition; instead, the ground state always stays at $\mathbf{Q} = 0$, without any transition to other $Q$ sectors.

Different from Fermi polarons in equal mass case, the $\mathbf{Q} = 0$ state exhibits rich inner structure associated with high-order correlations. To show this, we plot in Fig. 6a the respective weight of the bare term ($w_0$), one ($w_1$), two ($w_2$), and three ($w_3$) p-h excitation terms for $^{40}$K-$^{6}$Li Fermi polaron with $\eta = 40/6$. Specifically, the weight of $n$ p-h excitations terms is defined as

$$w_n = \frac{1}{(n!)^2} \sum_{\{\mathbf{k}\}\{\mathbf{q}\}} \left| \psi^{(n)}_{\mathbf{k}_1...\mathbf{k}_n \mathbf{q}_1...\mathbf{q}_n} \right|^2, \quad (8)$$

with the constraint $\sum_n w_n = 1$ from the normalization. One can see that as tuning the coupling $\ln(k_F a_{2d})$ from weak to strong, the system evolves from a polaronic state where $w_0$ dominates, to $w_1$-dominated and $w_2$-dominated intermediate states, and finally ends up at $w_3$-dominated state. This directly shows the smooth crossover in the $\mathbf{Q} = 0$ sector.

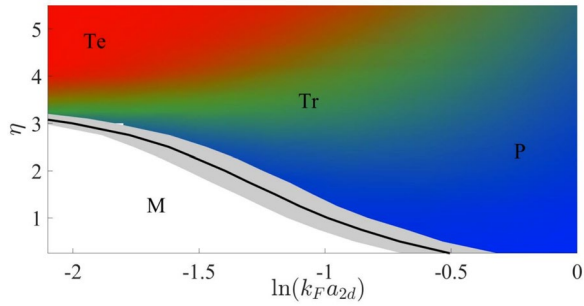

**Fig. 5** Phase diagram of the 2D Fermi polaron with interaction strength $\ln(k_F a_{2d})$ and mass ratio $\eta \equiv m_f/m_a$. The black line refers to the polaron-molecule phase boundary for small $\eta (< \eta_{tr} = 3.34)$. The gray area around the line denotes the coexistence region for polaron and molecule. For $\eta > \eta_{tr}$, no phase transition is found; Instead, as increasing the attraction strength, the system undergoes a smooth polaron-trimer crossover for $\eta \in (\eta_{tr}, \eta_{te})$ and a polaron-trimer-tetramer crossover for $\eta > \eta_{te} = 3.38$. The RGB color map is provided for the corresponding weights in the full wave-function of $Q = 0$ state, with $w_3$, $w_2$ and $w_1 + w_0$ representing the mixing ratio of R (*red*), G (*green*) and B (*blue*) colors, respectively. Adapted from Ref. [36]

Importantly, in the $w_2$-dominated and $w_3$-dominated regimes, the p-h excitations are strongly correlated near the Fermi surface. To extract such correlation, one can compute the hole-hole and particle-particle correlation functions of majority fermions in momentum space, which are defined as

$$D_h(\mathbf{q}_0, \mathbf{q}) \equiv \left\langle \left(1 - n^f_{\mathbf{q}_0}\right)\left(1 - n^f_{\mathbf{q}}\right) \right\rangle, \quad (9)$$

$$D_p(\mathbf{k}_0, \mathbf{k}) \equiv \left\langle n^f_{\mathbf{k}_0} n^f_{\mathbf{k}} \right\rangle, \quad (10)$$

with all $\mathbf{q}$ ($\mathbf{k}$) staying below (above) the Fermi surface. In Fig. 6b1–b3, we plot $D_h$ and $D_p$ together in momentum space (as varying $\mathbf{q}$ and $\mathbf{k}$), while keeping $\mathbf{q}_0$ and $\mathbf{k}_0$ fixed nearby the Fermi surface. In the polaron regime (b1), both $D_h$ and $D_p$ are extensively distributed in a broad angular range near the Fermi surface. In comparison, the correlation develops a visible crystalline feature in the $w_2$-dominated and $w_3$-dominated regimes, which respectively shows a diagonal (b2) and triangular (b3) structure. They are the same crystalline pattens as shown in the two-body correlations of trimer and tetramer bound states in vacuum [74]. Therefore the $w_2$- and $w_3$-dominated states can be seen, respectively, as the trimer and tetramer states dressed by particle-hole excitation of majority fermions.

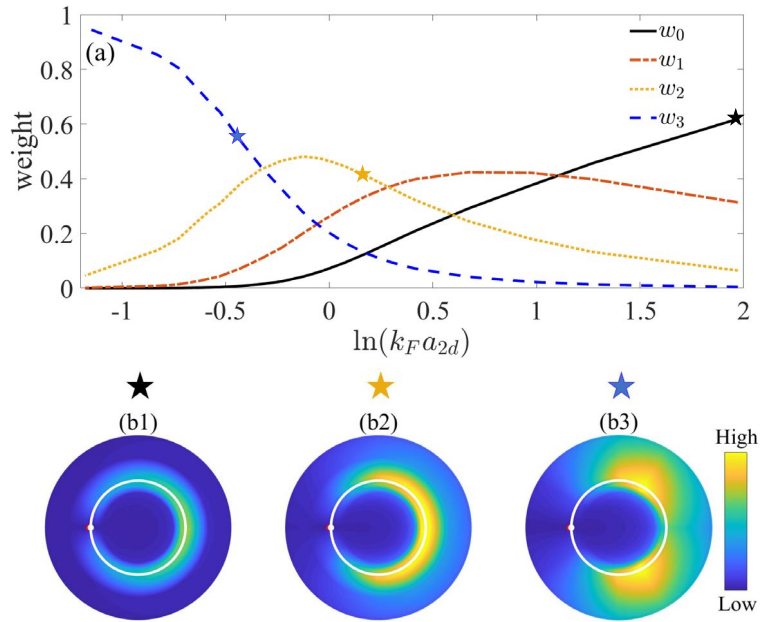

**Fig. 6** Polaron-trimer-tetramer crossover at mass ratio $\eta = 40/6$. **a** Weights of the bare term and various p-h excitation terms (see the $w_n$ definition in (8)) in $P_7(0)$ as functions of $\ln(k_F a_{2d})$. **b1–b3** Contour plots of hole-hole correlation $D_h(q_0, q)$ and particle-particle correlation $D_p(k_0, k)$ for different $\ln(k_F a_{2d}) = 1.972$ (**b1**), 0.172 (**b2**), and −0.428 (**b3**), which, respectively, belong to the polaron, trimer and tetramer regime as labeled by different stars in **a**. Here, we take $q_0 = -k_F e_x$ and $k_0 = -1.08 k_F e_x$, as marked by white and red points in the plots. The white circle denotes the Fermi surface. Adapted from Ref. [36]

These results have demonstrated that as increasing the impurity-fermion attractions, the mass-imbalanced Fermi polarons undergo a smooth crossover from a polaronic state to dressed trimer and tetramer states. These results are in sharp contrast to the first-order transition reported in previous studies [24, 25, 31, 35], which were based on the separate treatments of polaron and trimer states. In this context, the unified ansatz used here has proved its own advantage in equally evaluating all different correlations in a single framework. In fact, the unified ansatz has been shown to produce considerably lower energies than the separated trimer or tetramer states [36].

In above two sections, we have shown different physical consequences brought by different few-body correlations. Namely, the dominated two-body correlation will lead to a first-order polaron-molecule transition (as discussed in mass-balanced case), while the dominated three- or four-body correlations lead to a smooth crossover from polaronic to dressed cluster states. In fact, this is related to the highly symmetric p-h excitations associated with high-order correlations, such as the diagonal or triangular patterns shown in Fig. 7b2, c2. When scattering with these highly symmetric p-h excitations of heavy fermions, the impurity always has the largest weight around $\mathbf{k}=0$ (see Fig. 7b1, c1), similar to the polaronic state in weak coupling limit. It is then natural to expect the conversion between these states is smooth crossover. However, in the molecule case, the impurity has the largest weight around $|\mathbf{k}|=k_F$ (see Fig. 7a) in sharp contrast to the polaronic case. In this case the polaron-molecule conversion is a first-order transition. This demonstrates the remarkable difference between two- and higher-body correlation effects in Fermi polaron system.

## 5 Summary and outlook

In summary, we have discussed various competing few-body correlations and their physical consequences in attractive Fermi polarons. A crucial factor here is the mass ratio between the majority fermions and the impurity. For small mass ratios, the two-body correlation dominates and drives a polaron to molecule transition as the impurity-fermion attraction increases. This has been examined under a unified variational approach, which shows that the nature of such transition is the energy competition between different momentum sectors ($\mathbf{Q}=0$ and $|\mathbf{Q}|=k_F$). The advantage of such approach is the identification of coexistence region between polarons and molecules, as well as the hidden degeneracy of molecule states. When the mass ratios exceed certain values, the sharp transition is replaced by a smooth crossover between polaronic and dressed cluster states, and the latter regimes are associated with dominated high-body correlations. These correlations uniquely determine the crystalline patterns of heavy fermions in momentum space, which cannot be found in systems with two-body correlations. All these results could be detected in current cold atoms experiments of mass-imbalanced Fermi-Fermi mixtures.

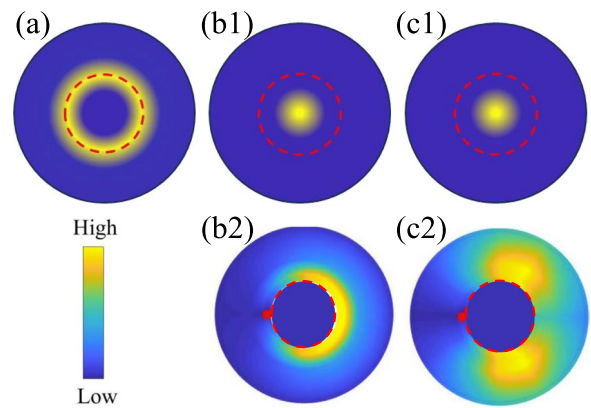

**Fig. 7** Schematics of momentum distribution/correlation of impurity and heavy fermions in various dressed states: molecule (**a**), trimer (**b1**, **b2**), and tetramer (**c1**, **c2**). The first panel show the single-particle distribution of impurity in molecule (**a**), trimer (**b1**), and tetramer (**c1**). The second panel shows the two-body correlation of heavy fermions in trimer (**b2**) and tetramer (**c2**) when fixing one fermion at the red point. The red (dashed) circle shows the Fermi surface of heavy fermions

It is interesting to explore in future the intriguing new phases brought by the high-order correlations in mass-imbalanced fermions. Along this direction, a recent work has proposed a novel high-order fermion superfluidity, i.e., the quartet superfluid, in the ground state of mass-imbalanced Fermi mixtures with heavy-light number ratio 3:1 [87]. Further studies of such intriguing superfluid include the finite temperature effect, phases with general polarizations, and the effects of finite-range or quasi-low dimensions as in realistic cold atoms experiments. Beyond the ultracold atomic systems, the mass-imbalanced fermion systems could also be realized in solid states where different effective masses can be associated with different bands, or in semiconductor TMD (transition metal dichalcogenides) systems where the mass-imbalanced mixture can consist of trions (charged excitons) and electrons (or holes) [88–92]. How the high-order correlations behave in these systems awaits to be explored in future.

## Acknowledgements
Not applicable.

## Authors' contributions
X.C. supervised the project and wrote the manuscript. R.L. collected data and provided the figures.

## Funding
The work is supported by the National Natural Science Foundation of China (12074419, 12134015), the Strategic Priority Research Program of Chinese Academy of Sciences (XDB33000000), the Fundamental Research Funds for the Central Universities (FRF-TP-24-040A), and the 2023 Fund for Fostering Young Scholars of the School of Mathematics and Physics, USTB (FRF-BR-23-01B).

## Data availability
Supporting information is available from the Springer website or from the authors.

## Declarations

## Competing interests
The authors declare that they have no competing interests.

Received: 12 August 2024   Accepted: 8 October 2024
Published online: 25 October 2024

## Publisher's Note